%% file: main.tex
\documentclass{sigplanconf}

\input{def/fonts}
\input{def/packages}
\input{def-import/macros}

\input{def/listings}
\input{def/macros}

\makeatletter
\def\@copyrightspace{\relax}
\makeatother

\begin{document}

\title{Behavioural Prototypes\vspace{-1ex}}
\authorinfo{Roly Perera\and Simon J.~Gay}{School of Computing Science,
  University of Glasgow, UK}{\{roly.perera, simon.gay\}@glasgow.ac.uk\vspace{-6ex}}

\maketitle

\input{body}

{\footnotesize \bibliographystyle{abbrv}
\bibliography{main.bib}}

\end{document}

%% file: def/fonts.tex
\edef\keptrmdefault{\rmdefault}
\edef\keptsfdefault{\sfdefault}
\edef\keptttdefault{\ttdefault}
\usepackage[math]{iwona}
\edef\rmdefault{\keptrmdefault}
\edef\sfdefault{\keptsfdefault}
\edef\ttdefault{\keptttdefault}

\usepackage[T1]{fontenc}
\usepackage{libertine}
\usepackage[scaled=0.75]{beramono}
\usepackage{upgreek}

%% file: def/packages.tex
\usepackage{amsmath}
\usepackage{breakurl}
\usepackage{calc} 
\usepackage[usenames,dvipsnames]{color}
\usepackage{etoolbox}
\usepackage{float}
\usepackage{hyperref}
\usepackage{listings}
\usepackage{microtype}
\usepackage{pifont} 
\usepackage[normalem]{ulem}
\usepackage{url}
\usepackage{xcolor}

%% file: def-import/macros.tex
\newcommand*{\kw}[1]{{\text{\tt{#1}}}} 

\DeclareSymbolFont{bbsymbol}{U}{bbold}{m}{n}
\DeclareMathSymbol{\bbsemi}{\mathbin}{bbsymbol}{"3B}

\newenvironment{nop}{}{}
\newenvironment{sdisplaymath}{
\begin{nop}\small\begin{displaymath}}{
\end{displaymath}\end{nop}\ignorespacesafterend}

\newenvironment{mathfig}{\begin{sdisplaymath}}{\end{sdisplaymath}}

\makeatletter
\newbox\sf@box
  {\def\sf@one{#1}%
   \def\sf@two{#2}%
   \setbox\sf@box\hbox
     \bgroup}%
  { \egroup
   \ifx\@empty\sf@two\@empty\relax
     \def\sf@two{\@empty}
   \fi
   \ifx\@empty\sf@one\@empty\relax
     \subfloat[\sf@two]{\box\sf@box}%
   \else
     \subfloat[\sf@one][\sf@two]{\box\sf@box}%
   \fi}
\makeatother

\definecolor{highlightcolor}{rgb}{1.0,0.8,0.8}
\definecolor{shadecolor}{rgb}{0.9,0.9,0.9}
\definecolor{lightgray}{rgb}{0.8,0.8,0.8}

\newsavebox{\vardisplaymathbox}

%% file: def/listings.tex
\definecolor{verylightgray}{gray}{0.9}
\definecolor{lightgray}{gray}{0.5}
\definecolor{mediumgray}{gray}{0.45}
\newlength\lsthorizontalpadding
\newcommand*\lstnumberstyle{\ttfamily\scriptsize\textcolor{lightgray}}
\newlength\lstnumbersep
\newlength\lstnumberwidth
\lstdefinelanguage{code}{%
   morekeywords={system,behaviour}%
  ,morecomment=[s]{\{-}{-\}}%
}
\lstnewenvironment{codecol}[1][]{\lstset{language=code,#1,moredelim=*[is][\textcolor{mediumgray}]{|}{|}}}{}

%% file: def/macros.tex
\makeatletter
\def\redwave{\bgroup \markoverwith{\lower3.5\p@\hbox{\sixly \textcolor{red}{\char58}}}\ULon}
\def\bluewave{\bgroup \markoverwith{\lower3.5\p@\hbox{\sixly \textcolor{blue}{\char58}}}\ULon}
\font\sixly=lasy6 
\makeatother

%% file: body.tex
\paragraph{Concurrent objects and multiparty compatibility.}

Data types describe values; behavioural types such as multiparty session
types \cite{honda08} and typestate \cite{garcia14} describe
interactions. Here we introduce a simple actor language and show
how \emph{multiparty compatibility} \cite{carbone15} can be used to
statically type-check systems of concurrent objects whose interfaces
evolve dynamically in response to messages.

Our program in our language is a collection of communicating
automata \cite{denielou13}. An asynchronous send of message \kw{m(v)}
to \kw{p} is written \kw{p!m(v)}. A blocking receive from \kw{p} is
written \kw{p?m(v)}, and binds the parameter
\kw{v} to the value sent. Our example models a simple software development workflow with
four mutually recursive roles. The wavy underlining can be ignored for
the moment.

In the code below, the \kw{teamLead} starts the \kw{devTeam} and then
begins a \kw{ReleaseCycle}. Once a release candidate at revision \kw{v}
is received from the \kw{devTeam}, the \kw{business} is informed. If the
\kw{business} accepts the release, it is tagged by the \kw{teamLead}
and the \kw{devTeam} can stop. Alternatively the \kw{business} can
request another iteration, and the cycle repeats.

\input{fig/example/teamLead}

The \kw{repository} enforces a source control protocol for the
organisation. Once a work unit has been committed by the \kw{devTeam},
the \kw{teamLead} must tag it appropriately before another work unit can
be accepted. The \kw{business} is a hard-coded test case that chooses
non-deterministically between \kw{iterate} and \kw{accept} for the first
release, and then accepts the second release without further ado.

\input{fig/example/repository-business}

The \kw{devTeam} commits work units to the \kw{repository},
simultaneously notifying the \kw{teamLead}, and bumping the revision
number at the end of each iteration. The interaction with the
\kw{math} library shown here is asynchronous; a more realistic language
would provide synchronous invocation as syntactic sugar.

\input{fig/example/devTeam}

If the \kw{commit} message to the \kw{repository} is omitted, as shown,
our implementation reports the compile-time errors shown as wavy lines.
Red underlining indicates that there is a
\kw{!} state of the system which is stuck because that message cannot
be delivered. Conversely, blue underlining indicates that there is a
\kw{?} state which is stuck because none of the permitted messages
arrives. Technically, these errors indicate \emph{multiparty
incompatibility}: the objects cannot be safely composed because a
coherent global session type cannot be derived which captures their
interactions. (A global session type certifies that the only stuck
states of a system are its terminal states \cite{carbone15}.) However,
note that we do not utilise a separate language of types: multiparty
compatibility is checked directly for objects.

\paragraph{Behavioural prototyping.}

For modularity, it is important to be able to define robust subsystem
boundaries. Our implementation does not support this yet, but in our
``typeless'' setting it would be natural for this role to be served by
concrete, executable objects, rather than interfaces, and for the notion
of implementation to be subsumed by behavioural subtyping. This would
permit declarations such as the following, which refines
the \kw{business} object. It defines a specialised implementation of
the \kw{releaseCandidate} handler which asks a \kw{customer} whether
to \kw{iterate} or \kw{accept}:

\input{fig/example/business-extended}

\noindent Intuitively, this is a valid refinement of the \kw{business} object because the \kw{customer}
interaction is private and preserves the observable behaviour
of \kw{business}. Implementation inheritance is mandatory.

\paragraph{Conclusion.}

We have shown some early features of our language and sketched an idea
for incorporating subtyping. Several non-trivial challenges lie ahead.
First, programs are not usually finite-state and so abstraction and
finitisation techniques will be required for multiparty compatibility to
remain decidable. Second, type errors reflect specific stuck states and
so to diagnose them properly may require integrating the type system
with a debugger. Finally, our language only supports systems with a
fixed set of roles, and will need extending to support dynamically
configured systems.

%% file: fig/example/teamLead.tex
{\small
\begin{lstlisting}
obj teamLead =
   devTeam!start;
   def ReleaseCycle =
      devTeam?releaseCandidate(v: number):
      business!releaseCandidate(v);
      business?
         iterate:
            $\smash{\texttt{\redwave{repository!tagRC(v)};}}$
            devTeam!continue;
            ReleaseCycle
         accept:
            $\smash{\texttt{\redwave{repository!tagRelease(v)};}}$
            devTeam!stop.
   ReleaseCycle
\end{lstlisting}
}

%% file: fig/example/repository-business.tex
{\small
\begin{lstlisting}
obj repository =
   def Connected =
      $\smash{\texttt{\bluewave{devTeam?commit(v: number):}}}$
      teamLead?
         tagRC(v: number): Connected
         tagRelease(v).
   Connected

obj business =
   teamLead?releaseCandidate(v: number):
   teamLead!
      iterate;
         teamLead?releaseCandidate(v: number):
         teamLead!accept.
      accept.
\end{lstlisting}
}

%% file: fig/example/devTeam.tex
{\small
\begin{lstlisting}
obj devTeam =
   teamLead?start:
   def ReleaseCycle(v: number) =
      $\texttt{\textcolor{lightgray}{// repository!commit(v);}}$
      teamLead!releaseCandidate(v);
      $\smash{\texttt{\bluewave{teamLead?}}}$
         $\smash{\texttt{\bluewave{continue:}}}$
            math!plus(v, 1); math?val(w: number):
            ReleaseCycle(w)
         $\smash{\texttt{\bluewave{stop}}}$.
   ReleaseCycle(0)
\end{lstlisting}
}

%% file: fig/example/business-extended.tex
{\small
\begin{lstlisting}
obj business': business =
   teamLead?releaseCandidate(v: number):
   customer!evaluate(v);
   customer?
      reject(comments: string):
         teamLead!iterate;
            teamLead?releaseCandidate(v: number):
            teamLead!accept.
      ok: teamLead!accept.
\end{lstlisting}
}